\documentstyle[prl,aps,twocolumn,epsf]{revtex}

\newcommand{\r}{{\bf r}}

\begin{document}
\twocolumn[\hsize\textwidth\columnwidth\hsize\csname
@twocolumnfalse\endcsname

\draft

\title{Quantum Monte Carlo Analysis of Exchange and Correlation in the 
Strongly Inhomogeneous Electron Gas}

\author{Maziar Nekovee,$^{1,*}$ W. M. C. Foulkes,$^1$ and R. J. Needs$^2$}
\address{$^1$The Blackett Laboratory, Imperial College, Prince 
Consort Road, London SW7 2BZ, UK \\$^2$Cavendish Laboratory,
Madingley Road, Cambridge CB3 0HE, UK}

\date{\today}

\maketitle

\begin{abstract}
\begin{quote}
\parbox{16 cm}{\small
We use variational quantum Monte Carlo to calculate the
density-functional exchange-correlation hole $n_{xc}$, the
exchange-correlation energy density $e_{xc}$, and the total
exchange-correlation energy $E_{xc}$, of several electron gas systems
in which strong density inhomogeneities are induced by a cosine-wave
potential.  We compare our results with the local density
approximation and the generalized gradient approximation.  It is found
that the nonlocal contributions to $e_{xc}$ contain an energetically
significant component, the magnitude, shape, and sign of which are
controlled by the Laplacian of the electron density.}
\end{quote}
\end{abstract}

]

\narrowtext

Kohn-Sham density-functional theory (DFT) \cite{KS} shows that it is
possible to calculate the ground-state properties of interacting
many-electron systems by solving only one-electron
Schr\"{o}dinger-like equations.  The results are exact in principle,
but in practice it is necessary to approximate the unknown exchange
and correlation (XC) energy functional, $E_{xc}[n(\r)]$, which
expresses the many-body effects in terms of the electron density
$n(\r)$.  The current popularity of density-functional methods in
condensed matter physics, quantum chemistry, and materials science
reflects the remarkable success of fairly simple approximate XC energy
functionals.

In the local density approximation (LDA), the XC hole $n_{xc}$ about
an electron at $\r$ is assumed to be the same as in a uniform electron
gas of fixed density $n = n(\r)$.  This works surprisingly well, even
though real solids and molecules have strongly inhomogeneous electron
densities.  Almost all attempts to go beyond the LDA have been based
on studies of XC in the weakly inhomogeneous electron gas.  The widely
used generalized gradient approximation (GGA) \cite{GGA1,GGA2,GGA3}
incorporates information about first-order gradient corrections to the
LDA, as obtained in the limit of slowly varying electron densities.
This is done in a controlled way such that, for example, $n_{xc}$
satisfies known sum rules \cite{GGA1}, but without any guidance from
the behavior of $n_{xc}$ in the strongly inhomogeneous regime.  This
may explain why current GGAs, although better than the LDA, are not
consistently able to deliver the very high accuracy of $\sim$0.1 eV
required to study many chemical reactions.  To provide a new direction
in the search for more accurate and reliable approximate functionals,
we have used variational quantum Monte Carlo (VMC)
\cite{VMC} to explore exchange and correlation in the strongly
inhomogeneous electron gas.

In this letter we calculate the central quantities in DFT: the XC hole
$n_{xc}$, the XC energy density $e_{xc}$, and the total XC energy
$E_{xc}$.  We study an electron gas of average density $n^0$
corresponding to $r_s$$=$2 and induce strong inhomogeneities by
applying a cosine-wave potential $V_{q} \cos({\bf q}.\r)$.  Fixing
$V_{q}$ at $2.08 \epsilon^0_F$, we consider several values of $q\leq
2.17 k_F^0$, where $\epsilon^0_F$ and $k_F^0$ are the Fermi energy and
wave vector of a uniform electron gas with $r_s$$=$2.

The electron density in our systems varies strongly and rapidly on the
scale of the inverse local Fermi wavevector
$k_F(\r)^{-1}=(3\pi^2n(\r))^{-1/3}$, producing a strikingly nonlocal
behavior of $n_{xc}$ that cannot be described by semilocal corrections
to the LDA.  We show, however, that the resulting LDA errors in
$e_{xc}$ have a dominant and energetically significant component, the
magnitude, shape, and sign of which are controlled by
$\nabla^{2}n(\r)$.  The GGA is unable to correct the LDA errors in
$E_{xc}$ resulting from this component in an adequate way, a
deficiency that severely restricts its accuracy.  The relevance of
Laplacian terms has been pointed out previously
\cite{Laplace,meta1,meta2}, but our calculations provide the first
quantitative evidence of their importance in strongly inhomogeneous
systems and for $e_{xc}$.

The starting point of our calculations is the adiabatic connection
formula \cite{parr}.  In Hartree atomic units
($e=m=4\pi\epsilon_0=\hbar=1$), this expresses $E_{xc}[n]$ as the
volume integral of an XC energy density $e_{xc}$ defined by:

\begin{equation}
\label{eq:exc}
e_{xc}(\r,[n]) = \frac{1}{2}\int d\r' \; 
\frac{n(\r)n_{xc}(\r,\r')}{|\r -\r'|} \;,
\end{equation}
\noindent
where $n_{xc}$ is the coupling-constant-averaged XC hole.  The
definition of $e_{xc}$ used in the construction of the GGA differs
from ours by an integration by parts that does not affect the
integrated $E_{xc}$.  We favor Eq.~(\ref{eq:exc}), however, because of
its clear physical interpretation and because it aids comparison with
the LDA XC energy density, which is constructed from an approximate
hole.

The XC hole $n_{xc}$ is obtained  via a coupling-constant integration,

\begin{eqnarray}
\label{eq:nxc}
\lefteqn{ n(\r)n(\r') + n(\r)n_{xc}(\r,\r') } \nonumber \\
& = &  \int_{0}^{1} d\lambda \;
\langle \Psi_{\lambda}| \sum_i \sum_{j (\neq i)}
\delta(\r-\r_i)\delta(\r'-\r_j) |
\Psi_{\lambda} \rangle \; ,
\label{eq:ccint}
\end{eqnarray}

\noindent
where $\Psi_{\lambda}$ is the antisymmetric ground state of the
Hamiltonian $\hat{H}^{\lambda}=\hat{T}+\lambda\hat{V}_{ee}
+\hat{V}^{\lambda}$ associated with coupling constant $\lambda$.  Here
$\hat{T}$ and $\hat{V}_{ee}$ are the operators for the kinetic and
electron-electron interaction energies, and 
$\hat{V}^{\lambda} = \Sigma V^{\lambda}(\r_i)$ is the
one-electron potential needed to hold the electron density
$n^{\lambda}(\r)$ associated with $\Psi_{\lambda}$ equal to $n(\r)$
for all values of $\lambda$ between 0 and 1.  At full coupling
($\lambda$$=$$1$), $V^{\lambda}(\r)$ coincides with the external
potential of the system.  For other $\lambda$ values $V^{\lambda}$
must be determined.

Our VMC method \cite{maziar1} for calculating $n_{xc}$ and $e_{xc}$
from Eqs.~(\ref{eq:exc}-\ref{eq:nxc}) is a generalization of the
scheme used by Hood {\it et al.}\ \cite{randy}.  It amounts to
treating {\em both} $\Psi^{\lambda}$ and $V^{\lambda}$ variationally
and determining the variational parameters by simultaneously
minimizing the variance of the local energy \cite{umrigar,andrew} and
the deviation of $n^{\lambda}$ from $n$ \cite{maziar1}.  This is
achieved by adding a term proportional to $|n^{\lambda}$-$n|$ to the
variance of the energy and minimizing the resulting penalty function,
which attains its absolute minimum of zero only when $\Psi^{\lambda}$
satisfies the Schr\"{o}dinger equation associated with $\lambda$
\emph{and} $n(\r) = n^{\lambda}(\r)$.

The VMC calculations are performed for a finite spin-unpolarized
electron gas in a FCC simulation cell subject to periodic boundary
conditions.  We generate the ground-state density by applying the
cosine-wave potential to the $\lambda$$=$$0$ system and solving the
self-consistent Kohn-Sham equations within the LDA to obtain the
single-particle orbitals $\phi_i$.  We then {\em define} the resulting
density to be the ground-state density of the fully interacting system
\cite{procedure}.  We consider potentials with $q = 1.11k_F^{0}$,
$1.55k_F^{0}$, and $2.17k_F^{0}$, and cells containing $64$, $78$, and
$69$ electrons, respectively.  The large amplitude of the cosine-wave
potential ensures that these systems are in the strongly inhomogeneous
regime, $|n(\r)-n^0|/ n^0 \approx 1$.

The adiabatic calculations were performed using $6$ equidistant values
of $\lambda$ in the range $[0,1]$, and with the following
Slater-Jastrow ansatz for $\Psi^{\lambda}$:

\begin{equation}
\Psi ^{\lambda} = 
D^{\uparrow }D^{\downarrow }\exp \left[ -\sum_{i>j}u^{\lambda}_{\sigma_i,
\sigma_j}(r_{ij})+\sum_i\chi^{\lambda} ({\bf r}_i)\right] \;,
\end{equation}

\noindent
where $r_{ij} = |\r_i - \r_j|$ and $D^{\uparrow }$ and $D^{\downarrow
}$ are spin-up and spin-down Slater determinants constructed using the
exact Kohn-Sham orbitals $\phi_i$.  The two-body term
$u^{\lambda}_{\sigma_i,\sigma_j}$ contained both a fixed part and a
variational part as discussed in \cite{maziar1}.  The one-body term
$\chi^{\lambda}$, the potential $V^{\lambda}$, and the electron
density $n^{\lambda}$ were all expanded in plane waves.  At each
$\lambda$ we used a total of $20$ variational parameters in
$u^{\lambda}$ and $\chi^{\lambda}$ and up to $7$ coefficients in the
plane-wave expansions of $n^{\lambda}$ and $v^{\lambda}$.  The
optimization of the parameters in $\Psi^{\lambda}$ and $V^{\lambda}$
was performed using $96000$ statistically uncorrelated electron
configurations.  This was sufficient to reduce the root mean square
deviation of $n^{\lambda}(\r)$ from $n(\r)$ to less than $0.5\%$ of
$n(\r)$ for all values of $\lambda$ and all systems.  After
optimization, expectation values were calculated using the Monte Carlo
Metropolis method \cite{VMC} with $10^6$ independent configurations of
all electrons.  Throughout we used the modified electron-electron
interaction described in \cite{matthew}, which virtually eliminates
the finite-size errors arising from the long range of the Coulomb
potential.  The statistical errors were negligible except in the tails
of $n_{xc}$ in low-density regions, where they were less than $3\%$;
this is much smaller than the differences between the VMC and LDA XC
holes.

The largest systematic errors are caused by the finite size of the
system and the approximate nature of $\Psi^{\lambda}$.  These errors
combine such that, even in a homogeneous electron gas, $e_{xc}^{VMC}
\neq e_{xc}^{LDA}$.  To circumvent this problem we performed
additional VMC calculations for finite homogeneous electron gases with
$N$$=$$64$ and $r_s = 0.8$, 1, 2, 3, 4, 5, 8, and 10.  The results
enabled us to construct a Perdew-Zunger (PZ) parameterization
\cite{PZ} of the VMC XC energy per electron of a finite uniform
electron gas with $N$$=$$64$.  This parameterization was used to
calculate $e_{xc}^{LDA}$ and $e_{xc}^{GGA}$ in all systems studied,
ensuring that these quantities were exactly equal to $e_{xc}^{VMC}$ in
any finite homogeneous system with $N$$=$$64$.  This procedure largely
eliminates the systematic errors in the calculated differences between
$e_{xc}^{VMC}$ and $e_{xc}^{LDA}$ reported below.

In Fig.~\ref{fig:holes} we show snapshots of the deformation of
$n_{xc}^{VMC}$ around an electron moving in the $q$$=$$1.55k_F$ system
along a line parallel to ${\bf q}$, the direction of maximum
inhomogeneity, from a density maximum towards the tail of $n(\r)$.
The XC hole is plotted as a function of $\r'$ around a fixed electron
at $\r$, with $\r'$ ranging in a plane parallel to ${\bf q}$.  Also
shown is the corresponding LDA hole $n_{xc}^{LDA}$ \cite{parr}.  At
the density maximum (not shown) both $n_{xc}^{VMC}$ and $n_{xc}^{LDA}$
are centered around the electron.  However, unlike $n_{xc}^{LDA}$,
which is always spherically symmetric, $n_{xc}^{VMC}$ is contracted in
the direction of inhomogeneity and less compact.  As the electron
moves away from the density maximum to a point on the slope (top
panel), the nonlocal nature of $n_{xc}^{VMC}$ becomes manifest.  While
$n_{xc}^{LDA}$ is still centered around the electron and is rather
diffuse, $n_{xc}^{VMC}$ lags behind near the density maximum and is
much more compact.  The nonlocal behavior of $n_{xc}^{VMC}$ becomes
remarkable at the density minimum.  Here $n_{xc}^{VMC}$ has two large
nonlocal minima, each centered around a density maximum $\sim$$2.80$
a.u.\ away from the electron.  The LDA hole at this point is much more
long ranged than $n_{xc}^{VMC}$ and is spread over the whole system in
order to satisfy the sum rule \cite{parr}: $\int d\r'
n_{xc}^{LDA}(\r,\r')=-1$.

This striking nonlocality of $n_{xc}^{VMC}$ also occurs in the other
two systems we considered.  Clearly, semilocal corrections are unable
to significantly improve the LDA description of the XC hole in our
systems, and fully nonlocal approximations are required.  We found,
however, that despite the strong nonlocality of $n_{xc}^{VMC}$, the
LDA errors in $e_{xc}^{VMC}$ \emph{can} be described in terms of a
semilocal quantity, the Laplacian of the electron density.
 
\begin{figure}
\begin{center}
\leavevmode
\epsfxsize=7cm \epsfbox{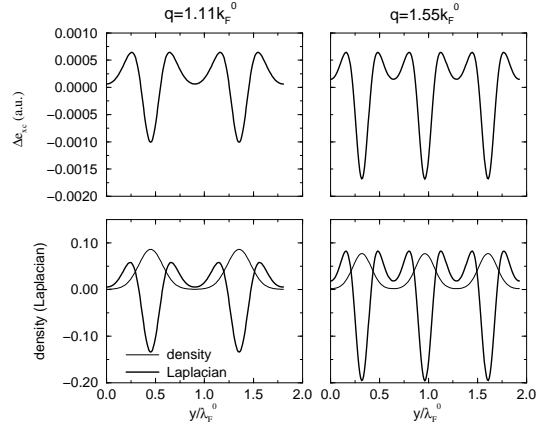}
\end{center}
\caption{The upper graphs show $e_{xc}^{LDA}-e_{xc}^{VMC}$ along a 
direction parallel to ${\bf q}$ for two different strongly
inhomogeneous systems. The lower graphs show the corresponding
electron densities (light lines) and Laplacians (heavy lines).
Distances are in units of the Fermi wavelength
$\lambda_F^0=2\pi/k_F^0$.  }
\label{fig:exc}
\end{figure}

In Fig.~\ref{fig:exc} we show $e_{xc}^{LDA}-e_{xc}^{VMC}$ for two of
the strongly inhomogeneous systems studied, where $e_{xc}^{LDA}$ is
calculated using the exact ground-state density $n(\r)$.  The results
are plotted along a line parallel to ${\bf q}$ (we call this direction
$y$).  Also shown are $n(\r)$ and $\nabla^{2}n(\r)$, plotted along the
same line.  It is apparent that the shape, magnitude, and sign of the
LDA errors in $e_{xc}$ closely follow the shape, magnitude, and sign
of $\nabla^{2}n(\r)$.  The LDA errors in $e_{xc}$ are large and
negative in regions where $\nabla^{2}n(\r)$ is large and negative
(around density maxima), and large and positive in regions where
$\nabla^{2}n(\r)$ is large and positive.  The GGA XC energy density is
not defined via Eq.~(\ref{eq:exc}) and so GGA errors are not shown.

The VMC values of the integrated $E_{xc}$ are shown in Table
\ref{tab:energies}, along with the differences $\Delta E_{xc}^{LDA} =
E_{xc}^{LDA} - E_{xc}^{VMC}$ and $\Delta E_{xc}^{GGA} = E_{xc}^{GGA} -
E_{xc}^{VMC}$  (The version of the GGA used here is due to Perdew,
Burke, and Ernzerhof \cite{GGA3}).  The LDA errors in $E_{xc}$ reflect
the profound effect of the Laplacian errors in $e_{xc}$ and change
sign from positive (for the $q$$=$$1.11k_F^{0}$ system ) to negative (for the
two other systems) as $q$ increases and the negative
contributions to $\Delta e_{xc}$, which occur where
$\nabla^{2}n(\r)<0$, become dominant.  The GGA corrections are by
construction {\em always} negative \cite{GGA1,GGA2,GGA3}; they improve
$E_{xc}^{LDA}$ for the $q$$=$$1.11 k_{F}$ system but worsen it for the
two other systems.

\begin{figure}
\begin{center}
\leavevmode
\epsfxsize=7cm \epsfbox{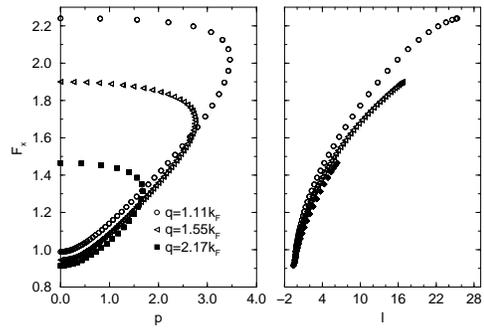}
\end{center}
\caption{
The left panel plots the exact enhancement factor $F_{x}$ against the
reduced density gradient for three strongly inhomogeneous systems. The
right panel shows the same quantity plotted against the reduced
density Laplacian.}
\label{fig:enhance}
\end{figure}

Since any smooth density $n(\r)$ may be expanded as a power series,
the XC energy density functional may always be written as
\begin{equation}
\label{eq:func}
e_{xc}(\r,[n]) = e_{xc}(\r, n(\r), \nabla_i n(\r), \nabla_i\nabla_j
n(\r), \ldots) \;.
\end{equation}
\noindent
The GGA may be viewed as an attempt to approximate the right-hand side
of Eq.~(\ref{eq:func}) as a nonlinear function of $n(\r)$ and
$\nabla_i n(\r)$.  In the case of the exchange energy, a uniform
scaling argument \cite{scaling} shows that
\begin{equation}
e_x = F_{x}(p,l,\ldots) e_{x}^{LDA} \;,
\end{equation}
\noindent
where $F_x$ is an enhancement factor, $e_x^{LDA}$ is the LDA exchange
energy density, and $p=|\nabla n|/(2k_F(\r)n(\r))$ and $l=\nabla^2
n/(4k_F^2(\r)n(\r))$ are a dimensionless gradient and a dimensionless
Laplacian, respectively.  We have calculated $F_x$ exactly for our
systems and investigated its dependence on $p$ and $l$.
Figs.~\ref{fig:enhance}a and \ref{fig:enhance}b are scatter plots
showing values of $F_x$ plotted against values of $p$ and $l$,
respectively.

The two-valued nature of Fig.~\ref{fig:enhance}a arises because the
mapping from density gradient to position is two valued in our
systems.  For example, the density gradient is zero both at the
minimum and the maximum of the density, where the required corrections
to the LDA XC hole are completely different, as can be inferred from
Fig.~\ref{fig:holes}.  In sharp contrast, $F_x$ is a simple and almost
unique function of $l$.  This suggests that the inclusion of Laplacian
terms may allow the construction of simpler and more accurate
approximate functionals.  In most gradient expansions the Laplacian
terms allowed by symmetry are transformed into $|\nabla n(\r)|^2$
terms via an integration by parts.  This is only possible when the
dependence on $l$ is linear, however, which is not the case here as
can be seen from Fig.~\ref{fig:enhance}b.  Furthermore, the
integration by parts destroys the physical interpretation in terms of
the XC hole and so hinders further progress.

We note that there is a one-to-one relationship between position and
$l$ in our structures, and so an enhancement factor of the form
$F_x(l)$ might not be as universal as Fig.~\ref{fig:enhance}b
suggests.  However, the energetic significance of the Laplacian terms,
and the strong similarity between the form of the Laplacian and the
LDA errors in $e_{xc}$ (see Fig.~\ref{fig:exc}), give us confidence in
the physical importance of the Laplacian in describing inhomogeneity
corrections to the LDA.  Very recently, Becke \cite{meta1} and Perdew
{\it et al.}\ \cite{meta2} have constructed beyond-GGA functionals
that contain the reduced Laplacian as a parameter.  We consider this a
step in the right direction and believe that the detailed \emph{local}
information on $n_{xc}$ and $e_{xc}$ in the strongly inhomogeneous
regime provided by our VMC calculations can be used to improve such
functionals.

We thank R.Q.\ Hood for useful discussions, G.\ Rajagopal and A.J.\
Williamson for help with the VMC codes, and L.\ Smith for help with
visualization.  M.N.\ was supported by an EU Marie Curie Fellowship
under Grant ERBFMBICT961736.  Our calculations were performed on the
CRAY-T3E at EPCC.

\begin{table}
\caption{Exchange-correlation energies (Hartrees per electron) and the
LDA and GGA errors in this quantity for different values of the wave
vector ${\bf q}$.  The statistical errors in $E_{xc}^{VMC}$ are indicated.}
\label{tab:energies}
\begin{tabular}{llll}
\hline
\\
 $q/k_F^0$ &$E_{xc}^{VMC}$ & $\Delta E_{xc}^{LDA}$ & $\Delta
 E_{xc}^{GGA}$ \\
\hline
1.11 & $-0.3289 \pm  0.0001$ & $+0.0042$   & $+0.0001$    \\
1.55 & $-0.3127 \pm  0.0001$ & $-0.0005$   & $-0.0074$     \\
2.17 & $-0.2882 \pm  0.0001$ & $-0.0066$   & $-0.0140$      \\
\hline
\end{tabular}
\end{table}

\begin{figure}
\begin{tabular}{@{}ccl@{}}
\end{tabular}
\caption{(color) The VMC and LDA $n_{xc}(\r,\r')$ for the 
$q$$=$$1.55 k_F^{0}$ system plotted around an electron fixed at $\r$
(indicated by a white bullet) with $\r'$ ranging in a plane parallel
to ${\bf q}$: (top) electron on the slope; (bottom) electron at a density
minimum. The VMC hole is shown on the left and the LDA hole on the 
right. The charge density is also represented schematically.
The color coding is only for $n_{xc}$ and varies between 
$-0.0500$ (blue) and $-0.0025=$ (red).}
\label{fig:holes}
\end{figure}

\begin{references}
\bibitem{procedure}
This has the advantage that the $\phi_i$ are by construction the exact
Kohn-Sham orbitals for density $n(\r)$, and their Slater determinant
is the exact wave function at $\lambda$$=$$0$.
\end{references}
\end{document}